\documentclass[a4paper,11pt]{article}
\usepackage{pos}
\usepackage{soul}
\usepackage{graphicx}

\pdfoutput=1

\title{Recent observations of PKS 2155-304 with MAGIC and LST-1 in a multi-wavelength context}
 \ShortTitle{PKS 2155-304 with MAGIC and LST-1}

\author*[a]{L.~Nikoli\'c}
\author[a]{G.~Verna}
\author[b]{M.~Manganaro}
\author[c]{G.~Bonnoli}
\author[f]{I.~Agudo}
\author[d]{G.~Silvestri}
\author[e]{D.~Cerasole}
\author[e]{F.~Schiavone}
\author[a]{F.~Podobnik}
\author[d]{J.~Otero-Santos}

\onbehalf{on behalf of the MAGIC collaboration and the CTAO-LST Project}

\affiliation[a]{INFN Sezione di Pisa and Università degli Studi di Siena, Dipartimento di Scienze Fisiche, della Terra e dell'Ambiente (DSFTA), Sezione di Fisica, Via Roma 56, 53100 Siena, Italy}

\affiliation[b]{Faculty of Physics, University of Rijeka, Radmile Matje\v{c}i\'c 2, Rijeka, 51000, Croatia}

\affiliation[c]{INAF - Osservatorio Astronomico di Brera, Via Brera 28, 20121 Milano, Italy}

\affiliation[d]{INFN Sezione di Padova and Università degli Studi di Padova, Via Marzolo 8, 35131 Padova, Italy}

\affiliation[e]{INFN Sezione di Bari and Università di Bari, via Orabona 4, 70126 Bari, Italy}

\affiliation[f]{Instituto de Astrofísica de Andalucía-CSIC, Glorieta de la Astronomía s/n, 18008, Granada, Spain}

\emailAdd{l.nikolic@student.unisi.it}

\abstract{
PKS~2155-304 is a well-known high-frequency peaked BL Lac (HBL) at redshift z=0.116, which has been extensively studied across the electromagnetic spectrum due to its rapid and large-amplitude variability. Several violent outbursts in X-rays and $\gamma$-rays have been observed in the past, with intra-night variability in very-high-energy $\gamma$-rays (VHE; E > 100 GeV) detected down to the minute timescale. The alternation of quiescent and enhanced states, observed with a tentative quasi-periodicity of 1.74 ± 0.13 years in high-energy (HE; 100 MeV < E < 100 GeV) $\gamma$-rays, makes this source a key target also for ground-based $\gamma$-ray instruments and in particular for the Imaging Atmospheric Cherenkov Telescopes. Its brightness, proximity, and well-determined redshift make this $\gamma$-ray source a prime target for fundamental physics studies, including tests of Lorentz Invariance Violation (LIV), searches for axion-like particles (ALPs), and constraints on the distribution of the extragalactic background light (EBL).
In the last two years, PKS~2155-304 has been independently monitored by the Major Atmospheric Gamma-ray Imaging Cherenkov (MAGIC) telescopes and the first Large-Sized Telescope (LST-1) of the Cherenkov Telescope Array Observatory (CTAO) located at the Roque de Los Muchachos Observatory (La Palma, Spain). The observations were carried out at large zenith angles (LZA; ZA > 55°), and the VHE data have been complemented with simultaneous observations in HE $\gamma$-rays (\textit{Fermi}-LAT), X-rays (\textit{Swift}-XRT) and optical wavelengths (ASAS-SN).
}

\ConferenceLogo{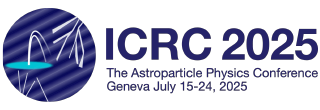}

\FullConference{39th International Cosmic Ray Conference (ICRC2025)\\
 15–24 July 2025\\
Geneva, Switzerland\\}

\begin{document}
\maketitle

\section{Introduction}

The active galactic nucleus (AGN) PKS~2155-304, located at a redshift of $z=0.116$ \cite{Falomo1993PKSredshift}, is one of the most luminous and extensively studied blazars belonging to the high-frequency peaked BL Lac (HBL), where the synchrotron peak frequency exceeds $ \nu_{peak} > 10^{15} \text{ Hz} $. These objects, characterized by relativistic jets closely aligned with the observer's line of sight, exhibit non-thermal emission that spans the entire electromagnetic spectrum, from radio to very-high-energy $\gamma$-rays (VHE; E > 100 GeV) \cite{UrryPadovani1995}. 
Its variability across all wavelengths and on timescales ranging from minutes to years makes it an interesting target for dedicated multi-wavelength (MWL) and multi-messenger campaigns.

PKS~2155-304 was recognized as an X-ray emitter by the HEAO-1 satellite in 1977 \cite{Schwartz1979HEAO1}.
The VHE emission of this source was detected for the first time by the University of Durham Mark VI Cherenkov telescope in 1997, making it the fourth X-ray-emitting BL Lac object confirmed as a VHE source and, at the time, the most distant TeV emitter detected from Earth \cite{Chadwick1999}.

The current generation of Imaging Atmospheric Cherenkov Telescopes (IACTs) such as High Energy Stereoscopic System (H.E.S.S.), the Major Atmospheric Gamma Imaging Cherenkov telescopes (MAGIC), and the Very Energetic Radiation Imaging Telescope Array System (VERITAS), along with continuous sky monitoring by the Large Area Telescope (LAT) on the \textit{Fermi Gamma-ray Space Telescope} (\textit{Fermi}) in the GeV band, have improved our understanding of PKS 2155-304. 
The source is known for its VHE variability, and perhaps the most significant example is the exceptional flare observed by H.E.S.S. in July 2006. During that flare, at energies above 200 GeV, the flux reached levels more than 10 times the quiescent emission ($\sim 7$ times the Crab Nebula flux, in the same energy range) and showed flux variations on timescales as short as 200 seconds \cite{Aharonian2007ApJL}.
Such rapid variability implies extremely compact emission regions and high Doppler factors, challenging simple one-zone emission models.
The same flare has made PKS 2155-304 a benchmark for testing CTAO's ability to track rapid variability, and that event is used to simulate and compare different emission scenarios \cite{consortium_science_2019}.
Long-term monitoring campaigns by H.E.S.S. \cite{HESS2010PKSlongterm} and observations by MAGIC \cite{Aleksic2012MAGIC} have provided extensive datasets, revealing complex flux and spectral changes, and have been important in characterizing its typical emission states and duty cycle. \textit{Fermi}-LAT observations have further covered its GeV emission, revealing periods of high activity and spectral evolution \cite{Abdo2010FermiPKS2155}.

The broadband spectral energy distribution (SED) of PKS~2155-304 is classically double-humped. The low-energy hump, peaking in the UV/X-ray range, is attributed to synchrotron radiation from relativistic electrons in the jet. The high-energy hump, extending to TeV energies, is generally interpreted as inverse Compton scattering of these same electrons, either off their own synchrotron photons (Synchrotron Self-Compton, SSC models) or off external photon fields (External Compton, EC models) \cite{Jagan2021PKS}.
The precise shape and variability of these components provide diagnostics of the underlying particle population and jet parameters. Coordinated MWL campaigns are crucial for understanding emission processes and have shown correlated variability between the X-ray and VHE bands, supporting leptonic models for many observed states, such as quiescent and flaring periods, characterized by different flux levels and spectral properties \cite{Katarzynski2005PKS}.

Beyond its role in blazar astrophysics, PKS~2155-304's bright VHE emission and well-determined redshift have made it a valuable source for recent studies of fundamental physics. The absorption of TeV photons by extragalactic background light (EBL) imprints a characteristic cutoff in its observed spectrum, allowing constraints on the EBL density \cite{collaboration_measurement_2013}.
Furthermore, rapid flares have been used to set limits on Lorentz Invariance Violation (LIV) by searching for energy-dependent photon propagation delays \cite{Abramowski2011HESSLIV}.

The MAGIC telescopes are located at the Observatorio del Roque de los Muchachos on La Palma (Canary Islands, Spain) at 2200\,m altitude. MAGIC consists of two 17\,m diameter IACTs separated by 85\,m, operating in stereoscopic mode to detect VHE $\gamma$-rays in the energy range from 50\,GeV up to 50\,TeV~\cite{Aleksic2015MAGIC}. 
The Cherenkov Telescope Array Observatory (CTAO) represents the next generation in ground-based VHE $\gamma$-ray astronomy, promising significantly improved sensitivity, angular resolution, and wider energy coverage compared to current instruments \cite{CTAConcept2011}. The first Large-Sized Telescope (LST-1) of the CTAO is situated about 100\,m away from MAGIC at the same site. LST-1 has a 23\,m diameter reflector and is optimized for detecting $\gamma$-rays from 20\,GeV to 200\,GeV, representing the first of four LSTs planned for the CTAO Northern array that are currently under construction. LST-1 has been operational since 2018 and is performing regular observations of various astrophysical sources, offering new opportunities to study objects like PKS 2155-304, particularly for capturing its low-energy VHE spectral features and rapid variability with high precision. Despite PKS~2155-304 having better visibility from the Southern Hemisphere, it is observable at large zenith angles (LZA; ZA > 55°) also from the MAGIC and LST-1 sites, and both collaborations have developed adequate analysis techniques to effectively perform LZA observations \cite{Peresano2018LZA}.

This contribution focuses on recent observations of PKS~2155-304 conducted with the MAGIC telescopes and the LST-1. By combining the capabilities of MAGIC and LST-1, particularly their higher sensitivity at lower energies, we aim to provide additional insights into the source's VHE emission characteristics. These observations are placed within a broader MWL context, incorporating data from other observatories, such as \textit{Fermi}-LAT, \textit{Swift}’s X-ray Telescope (XRT) and Ultraviolet/Optical Telescope (UVOT), and All-Sky Automated Survey for Supernovae (ASAS-SN). We present preliminary results, and discuss their implications for a better understanding of PKS~2155-304.


\section{Multi-wavelength Observations}

We are considering MWL observations of PKS~2155-304 between MJD~60000 and MJD~60700 (from February 25, 2023 to January 25, 2025) from optical to VHE energy band.
The observational details for each instrument are summarized in Table \ref{tab:observations}.
In the VHE band, LST-1 covers two different periods that are referred to in this work as P1 and P2.

\begin{table*}[ht]
\centering
\caption{Multi-wavelength observations of PKS 2155-304 in the period under study.}
\label{tab:observations}
\begin{tabular}{|l|c|c|c|c|}
\hline
\textbf{Instrument} & \textbf{Energy Band} & \textbf{Obs.\ Period (MJD)} & \textbf{Obs.\ Period (ISO 8601)} & \textbf{Obs.\ Time} \\
\hline
\textbf{MAGIC} & VHE & 60197--60259 & 2023-09-09 to 2023-11-11 & 9.4\,h$^a$ \\
\hline
\textbf{LST-1 (P1)} & VHE & 60196--60208 & 2023-09-08 to 2023-09-21 & 3.2\,h \\
\hline
\textbf{LST-1 (P2)} & VHE & 60460--60563 & 2024-05-30 to 2024-09-10 & 13.5\,h \\
\hline
\textbf{Fermi-LAT} & HE & 60000--60700 & 2023-02-25 to 2025-01-25 & --$^b$ \\
\hline
\textbf{Swift/XRT} & X-rays & 60000--60700 & 2023-02-25 to 2025-01-25 & -- \\
\hline
\textbf{Swift/UVOT} & UV/optical & 60000--60700 & 2023-02-25 to 2025-01-25 & -- \\
\hline
\textbf{ASAS-SN} & Optical & 60000--60700 & 2023-02-25 to 2025-01-25 & -- \\
\hline
\multicolumn{5}{l}{$^a$ One night discarded after quality cuts} \\
\multicolumn{5}{l}{$^b$ ``--'' indicates no discrete observing nights} \\
\end{tabular}
\end{table*}

During VHE $\gamma$-ray observations, weather conditions were monitored with the MAGIC LIDAR \cite{schmuckermaier_correcting_2023} and LST-1 weather station, respectively. MAGIC data were analyzed using dedicated MAGIC software \texttt{MARS} [v3.2.0] \cite{Moralejo2009MARS} and LST-1 data were analyzed using \texttt{lstchain} [v0.10.17] \cite{lstchain_adass_2020} and \texttt{Gammapy} [v1.3] \cite{Donath2023Gammapy}. 
The light curve (LC) was computed using night-wise binning for each.
In the HE $\gamma$-ray band we used data taken by \textit{Fermi}-LAT \cite{Atwood2009FermiLAT}. The analysis was done using \texttt{easyFermi} [v2.0.13] \cite{menezes_easyfermi_2022} software and LC was computed for the whole period using five day binning. 
X-ray observations were performed using \textit{Swift}/XRT \cite{Burrows2005SwiftXRT}. The reduced XRT data were downloaded from the UK Swift Science Data Center \citep{Evans2009}, and the spectral analysis was performed using the \texttt{XSPEC} [v12.13.1] package through \texttt{PyXspec}.
Ultraviolet (UV) observations were taken using \textit{Swift}/UVOT \cite{Roming2005SwiftUVOT}.
The analysis of UVOT data was carried out using the {\fontfamily{cmtt}\selectfont uvotimsum} and {\fontfamily{cmtt}\selectfont uvotsource} tasks within the \texttt{HEAsoft} [6.32] package, along with the \textit{Swift}/UVOT Calibration Database (CALDB) files [v20240201].
In the optical band, we used data collected by the ASAS-SN program \cite{shappee_man_2014}. It provided us with optical Sloan \textit{g} photometry data down to $\sim$17\,mag over the whole period. 
ASAS-SN, \textit{Swift}/XRT, and \textit{Swift}/UVOT light curves are binned per observation.


\section{Multi-wavelength Results}
\begin{figure}
    \centering
    \includegraphics[width=1\linewidth]{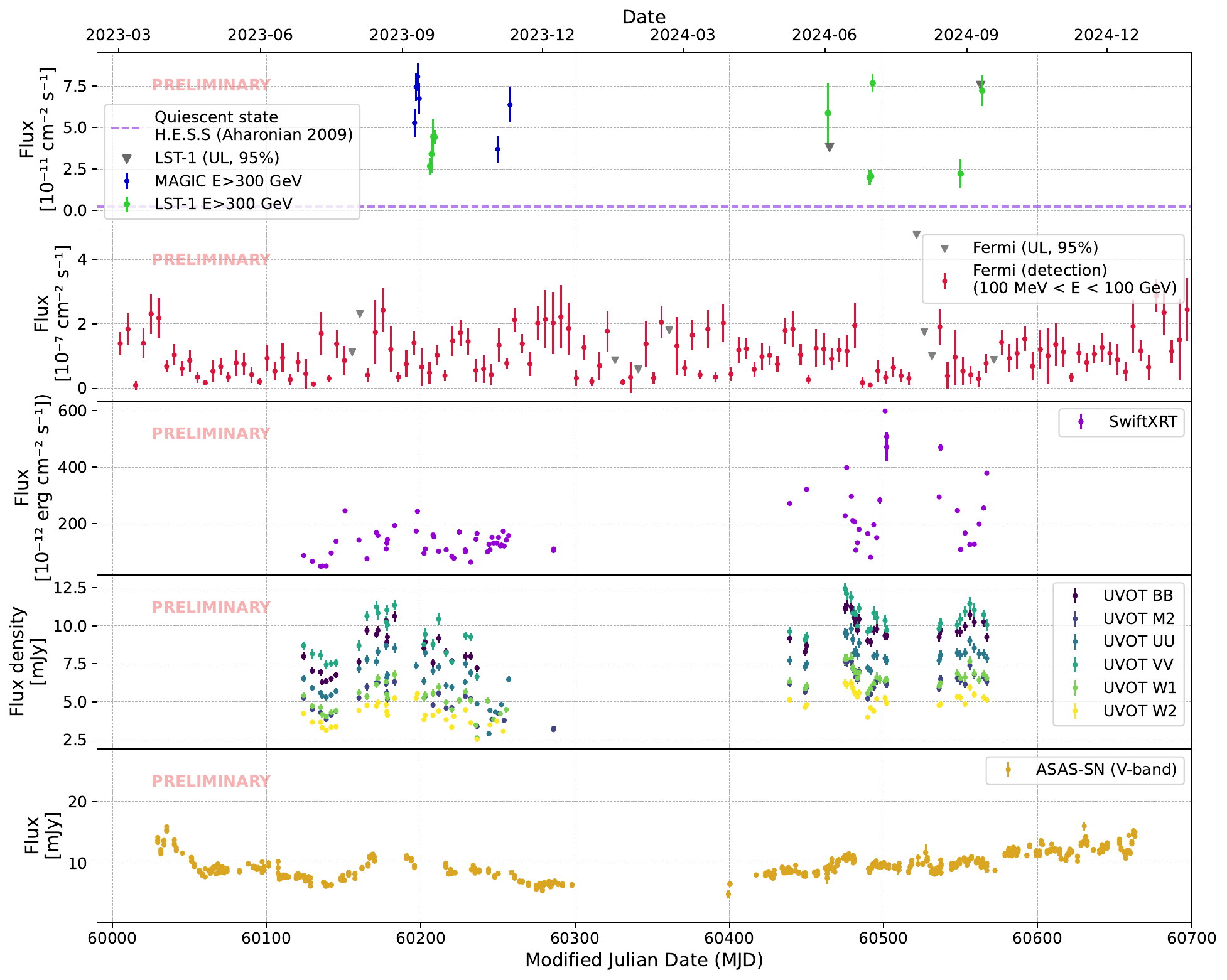}
    \caption{MWL Light Curves of the blazar PKS~2155-304 in the time range MJD~60000 and MJD~60700 (from February 25, 2023 to January 25, 2025). From top to bottom: MAGIC and LST-1, \textit{Fermi}-LAT, \textit{Swift}/XRT, \textit{Swift}/UVOT, and ASAS-SN).}
    \label{fig:MWL}
\end{figure}

In Fig.~\ref{fig:MWL} we show preliminary MWL light curves of PKS~2155-304 between MJD~60000 and MJD~60700, combining data from MAGIC (VHE $\gamma$-rays), LST-1 (VHE $\gamma$-rays), \textit{Fermi}-LAT (HE $\gamma$-rays), \textit{Swift}/XRT (X-rays), \textit{Swift}/UVOT (UV/optical), and ASAS-SN (optical \textit{g}-band). 
MAGIC detected a prominent VHE flare in the period from September to November 2023. To put this flare in context, we added a horizontal reference line to the figure showing typical quiescent flux levels measured by H.E.S.S. \cite{hess_simultaneous_2009}. The preliminary results of LST-1 suggest elevated VHE $\gamma$-ray states during both observation periods in 2023 and 2024, although the detailed analysis is still being finalized.
\textit{Fermi}-LAT shows no significant HE $\gamma$-ray flares throughout the entire observation period, maintaining relatively stable emission even during the VHE enhancement detected by MAGIC. \textit{Swift}/XRT and UVOT observations hint at enhanced activity in X-ray and UV/optical bands. During the VHE flare period, we see some X-ray activity, but the timing and amplitude relationships are more complicated than simple correlated variability would suggest. The UVOT data does not show clear extraordinary behavior during the X-ray peaks, and the Fermi flux was actually rather low during both the initial LST-1 period (P1) and the peak of XRT emission.
ASAS-SN monitoring shows long-term optical variability throughout the observation period, but these data require further detailed study to understand their link with the higher energy activity.

We have to emphasize that the apparent correlation is based on qualitative inspection of the light curves; a quantitative statistical correlation analysis is planned to better characterize the inter-band relationships and determine whether we are seeing correlated variability or more complex emission scenarios.

\section{Discussion And Summary}
Recent observations of PKS 2155-304 with the MAGIC and LST-1 telescopes, supported by comprehensive MWL data, provide new insights into the variability and emission mechanisms of this HBL. The qualitative simultaneity of flares in VHE, X-ray, and optical bands is suggestive of correlated variability and supports leptonic emission models.
The persistent HE emission and the complexity of inter-band correlations suggest that a scenario beyond the simple one-zone SSC models may be required, including the possibility that the HE $\gamma$-ray emission originates from a separate emitting region.
This scenario would be supported if the HE $\gamma$-ray variability shows different temporal behavior compared to the VHE, X-ray, and optical bands, indicating spatially distinct emission zones within the jet.
These results stress the value of joint observational campaigns and lay a solid foundation for future studies with CTAO and other next-generation instruments.

At this stage, we presented only preliminary temporal results. Spectral studies, including detailed modeling of the SED and cross-band correlations, are ongoing and will be discussed in future publications.

The presence of LST-1 alongside the MAGIC telescopes already enhances our observational capabilities, particularly in the lower end of the VHE range where LST-1 has increased sensitivity and lower energy threshold. While the observations presented here were taken independently by the two instruments, future joint observations will provide a more complete and simultaneous coverage of the VHE emission, enabling time-resolved spectral analyses better than the ones made with single-instrument datasets. Such coordinated efforts will improve cross-calibration, reduce systematic uncertainties, and allow for a deeper investigation of spectral features and variability of PKS~2155-304. These joint observations will be especially valuable for studying transitions between flux states and probing the underlying particle acceleration mechanisms with greater accuracy.

Looking ahead, completion of the four LSTs at the CTAO-North site will enable more sensitive and temporally resolved studies of sources like PKS~2155-304, especially during flaring states and at the low end of the VHE range.


\newpage

\section*{MAGIC Acknowledgements}
\tiny{
We would like to thank the Instituto de Astrof\'{\i}sica de Canarias for the excellent working conditions at the Observatorio del Roque de los Muchachos in La Palma. The financial support of the German BMBF, MPG and HGF; the Italian INFN and INAF; the Swiss National Fund SNF; the grants PID2019-107988GB-C22, PID2022-136828NB-C41, PID2022-137810NB-C22, PID2022-138172NB-C41, PID2022-138172NB-C42, PID2022-138172NB-C43, PID2022-139117NB-C41, PID2022-139117NB-C42, PID2022-139117NB-C43, PID2022-139117NB-C44, CNS2023-144504 funded by the Spanish MCIN/AEI/ 10.13039/501100011033 and "ERDF A way of making Europe; the Indian Department of Atomic Energy; the Japanese ICRR, the University of Tokyo, JSPS, and MEXT; the Bulgarian Ministry of Education and Science, National RI Roadmap Project DO1-400/18.12.2020 and the Academy of Finland grant nr. 320045 is gratefully acknowledged. This work was also been supported by Centros de Excelencia ``Severo Ochoa'' y Unidades ``Mar\'{\i}a de Maeztu'' program of the Spanish MCIN/AEI/ 10.13039/501100011033 (CEX2019-000920-S, CEX2019-000918-M, CEX2021-001131-S) and by the CERCA institution and grants 2021SGR00426 and 2021SGR00773 of the Generalitat de Catalunya; by the Croatian Science Foundation (HrZZ) Project IP-2022-10-4595 and the University of Rijeka Project uniri-prirod-18-48; by the Deutsche Forschungsgemeinschaft (SFB1491) and by the Lamarr-Institute for Machine Learning and Artificial Intelligence; by the Polish Ministry Of Education and Science grant No. 2021/WK/08; and by the Brazilian MCTIC, CNPq and FAPERJ.}

\section*{CTAO-LST Project Acknowledgments}
\tiny{
We gratefully acknowledge financial support from the following agencies and organisations:
Conselho Nacional de Desenvolvimento Cient\'{\i}fico e Tecnol\'{o}gico (CNPq), Funda\c{c}\~{a}o de Amparo \`{a} Pesquisa do Estado do Rio de Janeiro (FAPERJ), Funda\c{c}\~{a}o de Amparo \`{a} Pesquisa do Estado de S\~{a}o Paulo (FAPESP), Funda\c{c}\~{a}o de Apoio \`{a} Ci\^encia, Tecnologia e Inova\c{c}\~{a}o do Paran\'a - Funda\c{c}\~{a}o Arauc\'aria, Ministry of Science, Technology, Innovations and Communications (MCTIC), Brasil;
Ministry of Education and Science, National RI Roadmap Project DO1-153/28.08.2018, Bulgaria;
Croatian Science Foundation (HrZZ) Project IP-2022-10-4595, Rudjer Boskovic Institute, University of Osijek, University of Rijeka, University of Split, Faculty of Electrical Engineering, Mechanical Engineering and Naval Architecture, University of Zagreb, Faculty of Electrical Engineering and Computing, Croatia;
Ministry of Education, Youth and Sports, MEYS  LM2023047, EU/MEYS CZ.02.1.01/0.0/0.0/16\_013/0001403, CZ.02.1.01/0.0/0.0/18\_046/0016007, CZ.02.1.01/0.0/0.0/16\_019/0000754, CZ.02.01.01/00/22\_008/0004632 and CZ.02.01.01/00/23\_015/0008197 Czech Republic;
CNRS-IN2P3, the French Programme d’investissements d’avenir and the Enigmass Labex, 
This work has been done thanks to the facilities offered by the Univ. Savoie Mont Blanc - CNRS/IN2P3 MUST computing center, France;
Max Planck Society, German Bundesministerium f{\"u}r Bildung und Forschung (Verbundforschung / ErUM), Deutsche Forschungsgemeinschaft (SFBs 876 and 1491), Germany;
Istituto Nazionale di Astrofisica (INAF), Istituto Nazionale di Fisica Nucleare (INFN), Italian Ministry for University and Research (MUR), and the financial support from the European Union -- Next Generation EU under the project IR0000012 - CTA+ (CUP C53C22000430006), announcement N.3264 on 28/12/2021: ``Rafforzamento e creazione di IR nell’ambito del Piano Nazionale di Ripresa e Resilienza (PNRR)'';
ICRR, University of Tokyo, JSPS, MEXT, Japan;
JST SPRING - JPMJSP2108;
Narodowe Centrum Nauki, grant number 2023/50/A/ST9/00254, Poland;
The Spanish groups acknowledge the Spanish Ministry of Science and Innovation and the Spanish Research State Agency (AEI) through the government budget lines
PGE2022/28.06.000X.711.04,
28.06.000X.411.01 and 28.06.000X.711.04 of PGE 2023, 2024 and 2025,
and grants PID2019-104114RB-C31,  PID2019-107847RB-C44, PID2019-104114RB-C32, PID2019-105510GB-C31, PID2019-104114RB-C33, PID2019-107847RB-C43, PID2019-107847RB-C42, PID2019-107988GB-C22, PID2021-124581OB-I00, PID2021-125331NB-I00, PID2022-136828NB-C41, PID2022-137810NB-C22, PID2022-138172NB-C41, PID2022-138172NB-C42, PID2022-138172NB-C43, PID2022-139117NB-C41, PID2022-139117NB-C42, PID2022-139117NB-C43, PID2022-139117NB-C44, PID2022-136828NB-C42, PDC2023-145839-I00 funded by the Spanish MCIN/AEI/10.13039/501100011033 and “and by ERDF/EU and NextGenerationEU PRTR; the "Centro de Excelencia Severo Ochoa" program through grants no. CEX2019-000920-S, CEX2020-001007-S, CEX2021-001131-S; the "Unidad de Excelencia Mar\'ia de Maeztu" program through grants no. CEX2019-000918-M, CEX2020-001058-M; the "Ram\'on y Cajal" program through grants RYC2021-032991-I  funded by MICIN/AEI/10.13039/501100011033 and the European Union “NextGenerationEU”/PRTR and RYC2020-028639-I; the "Juan de la Cierva-Incorporaci\'on" program through grant no. IJC2019-040315-I and "Juan de la Cierva-formaci\'on"' through grant JDC2022-049705-I. They also acknowledge the "Atracci\'on de Talento" program of Comunidad de Madrid through grant no. 2019-T2/TIC-12900; the project "Tecnolog\'ias avanzadas para la exploraci\'on del universo y sus componentes" (PR47/21 TAU), funded by Comunidad de Madrid, by the Recovery, Transformation and Resilience Plan from the Spanish State, and by NextGenerationEU from the European Union through the Recovery and Resilience Facility; “MAD4SPACE: Desarrollo de tecnolog\'ias habilitadoras para estudios del espacio en la Comunidad de Madrid" (TEC-2024/TEC-182) project funded by Comunidad de Madrid; the La Caixa Banking Foundation, grant no. LCF/BQ/PI21/11830030; Junta de Andaluc\'ia under Plan Complementario de I+D+I (Ref. AST22\_0001) and Plan Andaluz de Investigaci\'on, Desarrollo e Innovaci\'on as research group FQM-322; Project ref. AST22\_00001\_9 with funding from NextGenerationEU funds; the “Ministerio de Ciencia, Innovaci\'on y Universidades”  and its “Plan de Recuperaci\'on, Transformaci\'on y Resiliencia”; “Consejer\'ia de Universidad, Investigaci\'on e Innovaci\'on” of the regional government of Andaluc\'ia and “Consejo Superior de Investigaciones Cient\'ificas”, Grant CNS2023-144504 funded by MICIU/AEI/10.13039/501100011033 and by the European Union NextGenerationEU/PRTR,  the European Union's Recovery and Resilience Facility-Next Generation, in the framework of the General Invitation of the Spanish Government’s public business entity Red.es to participate in talent attraction and retention programmes within Investment 4 of Component 19 of the Recovery, Transformation and Resilience Plan; Junta de Andaluc\'{\i}a under Plan Complementario de I+D+I (Ref. AST22\_00001), Plan Andaluz de Investigaci\'on, Desarrollo e Innovación (Ref. FQM-322). ``Programa Operativo de Crecimiento Inteligente" FEDER 2014-2020 (Ref.~ESFRI-2017-IAC-12), Ministerio de Ciencia e Innovaci\'on, 15\% co-financed by Consejer\'ia de Econom\'ia, Industria, Comercio y Conocimiento del Gobierno de Canarias; the "CERCA" program and the grants 2021SGR00426 and 2021SGR00679, all funded by the Generalitat de Catalunya; and the European Union's NextGenerationEU (PRTR-C17.I1). This research used the computing and storage resources provided by the Port d’Informaci\'o Cient\'ifica (PIC) data center.
State Secretariat for Education, Research and Innovation (SERI) and Swiss National Science Foundation (SNSF), Switzerland;
The research leading to these results has received funding from the European Union's Seventh Framework Programme (FP7/2007-2013) under grant agreements No~262053 and No~317446;
This project is receiving funding from the European Union's Horizon 2020 research and innovation programs under agreement No~676134;
ESCAPE - The European Science Cluster of Astronomy \& Particle Physics ESFRI Research Infrastructures has received funding from the European Union’s Horizon 2020 research and innovation programme under Grant Agreement no. 824064.}

\section*{\textbf{Full Author List: CTAO-LST Project}}

\tiny{\noindent
K.~Abe$^{1}$,
S.~Abe$^{2}$,
A.~Abhishek$^{3}$,
F.~Acero$^{4,5}$,
A.~Aguasca-Cabot$^{6}$,
I.~Agudo$^{7}$,
C.~Alispach$^{8}$,
D.~Ambrosino$^{9}$,
F.~Ambrosino$^{10}$,
L.~A.~Antonelli$^{10}$
% ...
The CTAO-LST Project
}\\

\tiny{\noindent$^{1}${Department of Physics, Tokai University, 4-1-1, Kita-Kaname, Hiratsuka, Kanagawa 259-1292, Japan}.
$^{2}${Institute for Cosmic Ray Research, University of Tokyo, 5-1-5, Kashiwa-no-ha, Kashiwa, Chiba 277-8582, Japan}.
$^{3}${INFN and Università degli Studi di Siena, Dipartimento di Scienze Fisiche, della Terra e dell'Ambiente (DSFTA), Sezione di Fisica, Via Roma 56, 53100 Siena, Italy}.
$^{4}${Université Paris-Saclay, Université Paris Cité, CEA, CNRS, AIM, F-91191 Gif-sur-Yvette Cedex, France}.
$^{5}${FSLAC IRL 2009, CNRS/IAC, La Laguna, Tenerife, Spain}.
$^{6}${Departament de Física Quàntica i Astrofísica, Institut de Ciències del Cosmos, Universitat de Barcelona, IEEC-UB, Martí i Franquès, 1, 08028, Barcelona, Spain}.
$^{7}${Instituto de Astrofísica de Andalucía-CSIC, Glorieta de la Astronomía s/n, 18008, Granada, Spain}.
$^{8}${Department of Astronomy, University of Geneva, Chemin d'Ecogia 16, CH-1290 Versoix, Switzerland}.
$^{9}${INFN Sezione di Napoli, Via Cintia, ed. G, 80126 Napoli, Italy}.
$^{10}${INAF - Osservatorio Astronomico di Roma, Via di Frascati 33, 00040, Monteporzio Catone, Italy}.
% ...
}

\end{document}